\documentclass[aps,prb,twocolumn,superscriptaddress,longbibliography]{revtex4-2}
\usepackage{epsfig,amsopn}
\usepackage{graphicx}
\usepackage{color}
\usepackage{amsmath,amssymb}
\usepackage{enumerate}
\newcommand\bea{\begin{eqnarray}}
\newcommand\eea{\end{eqnarray}}
\newcommand\beq{\begin{equation}}
\usepackage{gensymb}
\usepackage{verbatim}
\newcommand\eeq{\end{equation}}

\def\nn{\nonumber}

\def\de{\delta}

\def\ga{\gamma}

\def\si{\sigma}

\def\dg{\dagger}

\def\ua{\uparrow}
\def\da{\downarrow}

\def\th{\theta}

\begin{document}
\title{Equilibrium spin currents in altermagnet junctions: Josephson-like and anomalous transport} 
 \author{ Abhiram Soori}
 \email{abhirams@uohyd.ac.in}
 \affiliation{School of Physics, University of Hyderabad, Prof. C. R. Rao Road, Gachibowli, Hyderabad-500046, India}
\begin{abstract}
Altermagnets (AMs) offer a compelling platform for exploring novel spin-dependent phenomena in materials with zero net macroscopic magnetization. In this work, we theoretically investigate the emergence of equilibrium spin currents (ESCs) in two-dimensional AM heterostructures using a tight-binding lattice model. We first study an AM-normal metal-AM (AM-NM-AM) junction and demonstrate that the $\sigma_y$-polarized ESC exhibits a characteristic Josephson-like behavior, fundamentally governed by the relative angle ($\theta$) between the N\'eel vectors of the two AMs pointing in $xz$-plane. Crucially, we show that replacing the central normal metal with a $p$-wave magnet (PM) induces an anomalous ESC. Analogous to the anomalous Josephson effect, the breaking of spatial inversion symmetry by the PM allows a finite, dissipationless spin current to flow even when the N\'eel vectors are perfectly aligned ($\theta=0$). We establish that this anomalous transport is driven by an asymmetry in the quantum phases accumulated by right- and left-moving electrons undergoing  spin-flip reflections. Finally, we show that the critical ESC exhibits pronounced fluctuations as a function of band filling, which we attribute to mesoscopic quantum size effects, including transverse subband quantization and longitudinal Fabry-P\'erot resonances. Our findings highlight the potential of altermagnet junctions for designing dissipationless, phase-tunable spintronic devices.
\end{abstract}
\maketitle

\section{Introduction}

Ferromagnets, characterized by spin-polarized electrons, are broadly classified into metals and insulators. Nonequilibrium electron transport between two ferromagnetic metals yields fascinating effects, such as giant magnetoresistance~\cite{moodera1995}, which has driven major advancements in sensors and magnetoresistive random-access memory~\cite{akerman2005}. Conversely, ferromagnetic insulators are indispensable in spintronics and magnonics, where spin information is transported via magnons instead of conduction electrons.

Dissipationless equilibrium currents are fundamentally known to occur in two distinct physical arenas: mesoscopic normal metal rings and Josephson junctions. In mesoscopic rings, sustaining an equilibrium current necessitates the simultaneous breaking of both inversion and time-reversal symmetries~\cite{butti83,sahoo2025}. In contrast, a Josephson junction hosts an equilibrium supercurrent driven purely by the macroscopic superconducting phase bias between its two constituent superconductors~\cite{Josephson1962}.

Within the Bogoliubov–de Gennes (BdG) mean-field formalism, the superconducting pairing potential is proportional to $\tau_y$ in one superconductor and $(\cos{\phi_S}\tau_y + \sin{\phi_S}\tau_x)$ in the other. Here, $\phi_S$ denotes the superconducting phase difference, and $\tau_x$, $\tau_y$, and $\tau_z$ are the Pauli matrices spanning the particle-hole space. The corresponding charge density is defined as $\psi^{\dagger}\tau_z\psi$. The phase-driven Josephson current inherently satisfies the continuity equation alongside the charge density across the junction region.

By analogy, the Hamiltonian of a ferromagnet contains an exchange term proportional to $\sigma_z$, where $\sigma_z$ is the Pauli matrix representing the spin degree of freedom. By mapping the particle-hole symmetric space of a superconductor onto the two spin channels of a ferromagnet, a ferromagnet effectively behaves as the magnetic analogue of a superconductor. Therefore, joining two ferromagnets with non-collinear spin polarizations naturally induces the flow of an equilibrium spin current (ESC)~\cite{shen2007,chen2014}. This ESC is carried by the spin component that is strictly orthogonal to the local magnetization axes of both adjacent ferromagnets.

Recently, altermagnets (AMs) have emerged as a novel magnetic phase, unifying the hallmark features of both ferromagnets and antiferromagnets~\cite{smejkal22b,smejkal22c,fern24,yan24,reich24}. Conceptually parallel to $d$-wave superconductors, altermagnets host a strongly spin-split Fermi surface despite exhibiting strictly zero macroscopic net magnetization. In an AM, the spin projection along a designated axis remains a robust quantum number, parameterized by the so-called N\'eel vector. While macroscopic time-reversal symmetry ($\mathcal{T}$) is intrinsically broken, AMs are protected by a combined $C_4 \mathcal{T}$ symmetry, where $C_4$ represents a spatial $\pi/2$ rotation~\cite{ganesh2025}. Remarkably, even though neither a normal metal nor an AM possesses a net spin polarization, driving a bias across a junction between them can still generate a steady spin current~\cite{das2023}. 

Given that altermagnets act as the magnetic analogues of $d$-wave superconductors, it logically follows that an equilibrium spin current should emerge in a junction composed of two AMs with misaligned N\'eel vectors. In this work, we theoretically investigate this exact phenomenon within junctions comprising two AMs separated by a metal spacer. A critical distinction separates this ESC from the standard Josephson effect: whereas the Josephson supercurrent is mediated by discrete bound states localized within the superconducting energy gap, an altermagnetic metal possesses no such gap. Consequently, the ESC in these junctions is a bulk effect, carried collectively by all occupied plane-wave states.

Furthermore, we demonstrate that replacing the normal metal spacer with a $p$-wave magnet induces an anomalous ESC. Mirroring the anomalous Josephson effect~\cite{hasan2022}, the intrinsic breaking of inversion symmetry by the $p$-wave magnet permits a non-vanishing equilibrium spin current to flow even when the N\'eel vectors of the two adjacent AMs are perfectly collinear.

\begin{figure}
\includegraphics[width=8cm]{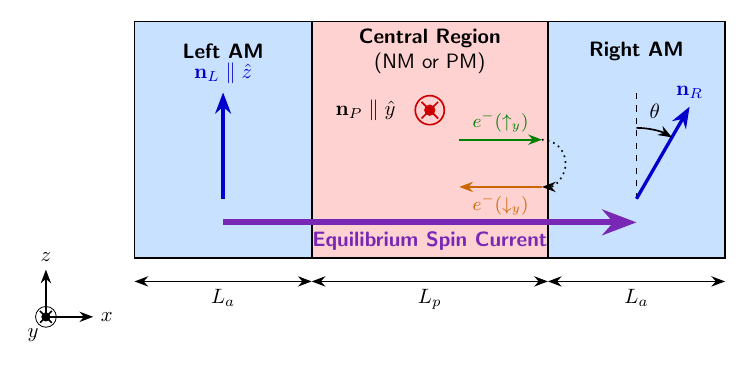}
\caption{Schematic diagram of the magnetic heterostructure. The setup consists of a central region (either an NM or a PM) sandwiched between two AMs with a relative N\'eel vector angle $\th$. This phase-like parameter $\th$ governs the flow of an equilibrium spin current through the junction, establishing a spintronic analog to the conventional Josephson effect.
The two AMs have their N\'eel vectors pointing in $xz$-plane, whereas the N\'eel vector of the central PM points along $y$-direction. }\label{fig:schem}
\end{figure}
\section{Details of calculation}
We consider an AM-PM-AM junction [see Fig.~\ref{fig:schem} for a schematic], where the central PM can be  tuned into a NM by setting the $p$-wave exchange strength to zero. The Hamiltonian for the system is given by $H=H_L+H_M+H_R+H_{LM}+H_{MR}$, where
\begin{widetext}
\bea 
H_L &=& 
-\sum\limits_{n_y=1}^{L_y}\sum\limits_{n_x=1}^{L_a-1}\big[ c^{\dg}_{n_x+1,n_y}(t\si_0+t_a\si_{\th})c_{n_x,n_y} + {\rm h.c.}\big]
-\sum\limits_{n_y=1}^{L_y-1}\sum\limits_{n_x=1}^{L_a}\big[ c^{\dg}_{n_x,n_y+1}(t\si_0-t_a\si_{\th})c_{n_x,n_y} + {\rm h.c.}\big], \nn \\ 
H_M &=&  -\sum\limits_{n_y=1}^{L_y}\sum\limits_{n_x=L_a+1}^{L_{ap}}\big[ c^{\dg}_{n_x+1,n_y}(t\si_0+i t_x\si_y)c_{n_x,n_y} + {\rm h.c.}\big]
-\sum\limits_{n_y=1}^{L_y-1}\sum\limits_{n_x=L_a+1}^{L_{ap}}\big[ c^{\dg}_{n_x,n_y+1}(t\si_0+i t_y\si_y)c_{n_x,n_y} + {\rm h.c.}\big], \nn \\ 
H_R &=& 
-\sum\limits_{n_y=1}^{L_y}\sum\limits_{n_x=L_{ap}+1}^{L_{apa}-1}\big[ c^{\dg}_{n_x+1,n_y}(t\si_0+t_a\si_{z})c_{n_x,n_y} + {\rm h.c.}\big]
-\sum\limits_{n_y=1}^{L_y-1}\sum\limits_{n_x=L_{ap}+1}^{L_{apa}}\big[ c^{\dg}_{n_x,n_y+1}(t\si_0-t_a\si_{z})c_{n_x,n_y} + {\rm h.c.}\big], \nn \\ 
H_{LM} &=& -t_l\sum\limits_{n_y=1}^{L_y}\big(c^{\dg}_{L_a+1,n_y}c_{L_a,n_y} +{\rm h.c.} \big), ~~~~H_{MR}~=~-t_r\sum\limits_{n_y=1}^{L_y}\big(c^{\dg}_{L_{ap}+1,n_y}c_{L_{ap},n_y}+{\rm h.c.}\big) .
\eea
\end{widetext}
Here,  $L_{ap}=L_a+L_p$ and $L_{apa}=2L_a+L_p$ specify the system dimensions. We define $\si_{\th}=(\cos\th\si_z+\sin\th\si_x)$, where $\si_j$ ($j=x,y,z$) are the Pauli spin matrices, $\si_0$ is the $2 \times 2$ identity matrix, $t$ represents the nearest-neighbor hopping amplitude within both the AM and PM regions, and $t_a$ characterizes the strength of the altermagnetic spin-splitting. The two-component spinor is denoted by $c_{n_x,n_y}=[c_{n_x,n_y,\ua},~c_{n_x,n_y,\da}]^T$, with $c_{n_x,n_y,\si}$ being the annihilation operator for an electron with spin $\si$ at site $(n_x,n_y)$. The $p$-wave magnetic terms are parameterized by $t_x=t_p\cos\phi$ and $t_y=t_p\sin\phi$, where $t_p$ defines the strength of the $p$-wave magnet. The angle $\phi$ specifies the orientation of the PM crystallographic axis along which the band bottoms for the two spin sectors are separated. Finally, $t_l$ ($t_r$) denotes the interfacial hopping amplitude that couples the left (right) AM to the central PM.

We diagonalize the full Hamiltonian and populate the energy eigenstates up to a specified filling fraction $f$. Since the spin operator $\si_y$ commutes with the Hamiltonian of the central PM region, the associated $y$-polarized spin current is conserved. The corresponding spin current operator across the interface is defined as $\hat J^s_y = -i\sum_{n_y}(c^{\dg}_{L_a+1,n_y}\si_yc_{L_s,n_y}-{\rm h.c.})$, where the transverse site index $n_y$ runs from $1$ to $L_y$. The total ESC is then evaluated by summing the expectation value of $\hat J^s_y$ over all occupied states.

\section{Results and Analysis}
\begin{figure}[htb]
\includegraphics[width=7cm]{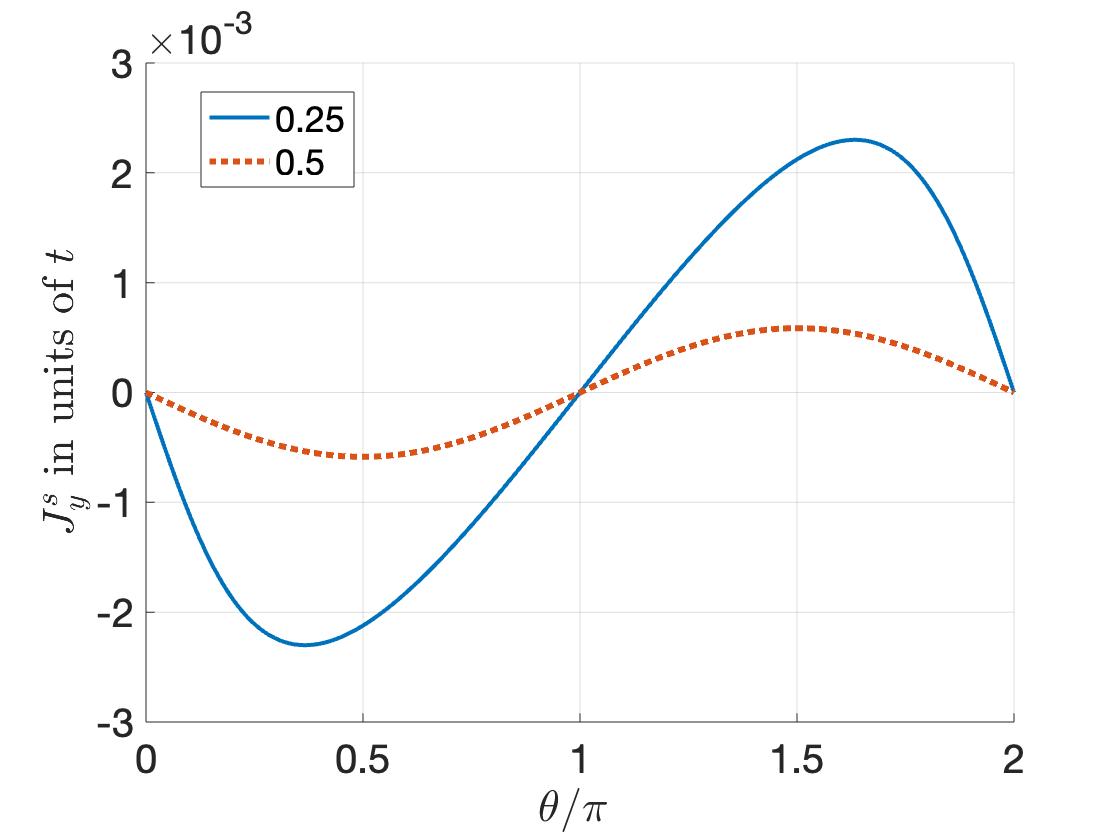}
\caption{Equilibrium spin current (ESC) as a function of the relative angle between the N\'eel vectors of the two AMs in an AM-NM-AM junction. The various curves correspond to different filling factors $f$, as indicated in the legend. System parameters are set to: $L_a=20$, $L_p=10$, $L_y=10$, $t_a=0.5$, $t_l=0.2t$, $t_r=0.1t$, and $t_p=0$.}\label{fig:cpr1}
\end{figure}

We first calculate the dependence of ESC on the relative angle between the N\'eel vectors of the two AMs in an AM-NM-AM junction. Because spin remains a good quantum number in the NM, the spin currents corresponding to $\si_x$, $\si_y$, and $\si_z$ are all well defined within the central region. We orient the N\'eel vector of the left AM along the $\hat z$-axis, while the N\'eel vector of the right AM is confined to the $\hat x-\hat z$ plane. Under this configuration, we find that only the ESC associated with $\si_y$ can be non-zero. In Fig.~\ref{fig:cpr1}, we plot this $\si_y$-polarized ESC, denoted as $J^y_x$, as a function of the relative angle between the N\'eel vectors. Similar to the behavior observed in conventional FM-NM-FM junctions, the ESC exhibits a characteristic Josephson-like effect.

\begin{figure}[htb]
\includegraphics[width=7cm]{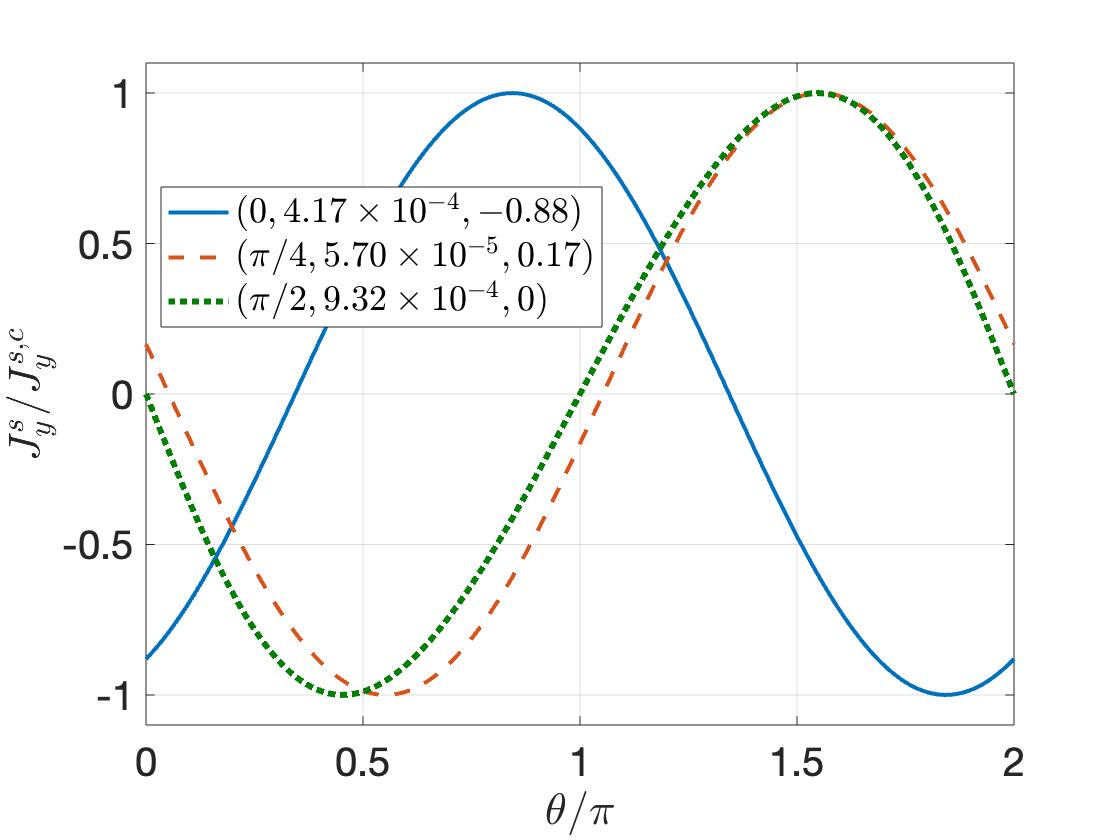}
\caption{ESC as a function of the relative angle $\th$ between the N\'eel vectors of the two AMs in an AM-PM-AM junction. Note the presence of a non-zero anomalous ESC, $J_y^{s,a}$, at $\th = 0$. The legend indicates the values of $(\phi, J_y^{s,c}, J_y^{s,a}/J_y^{s,c})$ for each respective curve. Note that anomalous ESC vanishes for $\phi=\pi/2$ due to restored inversion symmetry along $x$-direction.  System parameters are set to: $L_a=20$, $L_p=10$, $L_y=10$, $t_a=0.5t$, $t_l=0.2t$, $t_r=0.1t$,  $t_p=0.5t$, and $f=0.5$.}\label{fig:cpr-ano}
\end{figure}

Next, we replace the normal metal with a PM to study the ESC in an AM-PM-AM junction. Here, the N\'eel vector of the PM is oriented along $\hat y$, ensuring that $\si_y$ commutes with the central region's Hamiltonian and that the corresponding ESC remains well defined. Crucially, the Hamiltonian for the $p$-wave magnet breaks spatial inversion symmetry. Just as the simultaneous breaking of inversion and time-reversal symmetries is a prerequisite for the anomalous Josephson effect in superconducting junctions, this symmetry breaking allows the ESC to exhibit an anomalous spin current when the inversion-breaking PM is sandwiched between the AMs. As shown in Fig.~\ref{fig:cpr-ano}, which plots the ESC as a function of the relative angle $\th$, the anomalous nature is clearly evident: the ESC remains manifestly non-zero even at $\th=0$.

The dispersion relation for the central PM region is given by $E=-2t(\cos{k_x}+\cos{k_y})-(2t_x\sin{k_x}+2t_y\sin k_y)\si_y$. For a fixed transverse momentum $k_y$, this relation reveals a crucial asymmetry at a given energy when the N\'eel vectors of the two AMs are parallel: the phase $\phi_F$ accumulated by right-moving up-spin and left-moving down-spin electrons differs from the phase $\phi_B$ accumulated by left-moving up-spin and right-moving down-spin electrons. This phase difference fundamentally drives the anomalous ESC, analogous to the microscopic origin of the anomalous Josephson effect~\cite{soori2024bamjde}. To be more precise, the longitudinal momentum satisfies $k_x-\si_y\eta=\pm\cos^{-1}[f(\si_y,E)]$, where 
\[ 
f(\si_y,E) = \frac{E+2t\cos{k_y}+\si_y 2t_y\cos{k_y}}{2\sqrt{t^2+t_x^2}}, 
\] 
and the phase shift $\eta$ is defined by $(\cos\eta,\sin\eta)=(t,t_x)/\sqrt{t^2+t_x^2}$. Here, $\si_y$ takes the value $\pm 1$ for up and down spins, respectively, and the $\pm$ sign preceding the $\cos^{-1}$ term corresponds to right- and left-moving states. 
Furthermore, the phase of the reflection amplitude for an up-spin electron reflecting as a down-spin electron at the right interface takes the form $\th+\de$. Conversely, the reflection amplitude for a down-spin electron reflecting as an up-spin electron has the phase $-\th+\de$. 
For a complete back-and-forth traversal of a right-moving up-spin and a left-moving down-spin electron, the total accumulated phase is $\phi_F=(2\eta+\ga)L+\th+\de$, where $\ga=\cos^{-1}[f(+1,E)]+\cos^{-1}[f(-1,E)]$. Similarly, the total phase accumulated by a right-moving down-spin and a left-moving up-spin electron is $\phi_B=(-2\eta+\ga)L-\th+\de$. 
For a given angle $\th$, replacing $\th$ with $-(4\eta L+\th)$ precisely interchanges the entries of the phase pair $(\phi_F,\phi_B)$. Physically, this implies that for every forward-moving spin current contribution, there exists a backward-moving spin current of equal magnitude. This symmetry is the underlying reason for the absence of a nonreciprocal ESC. 
Finally, in the normal metal limit ($t_p=0$), the phase shift $\eta$ vanishes. In this regime, for a given nonzero relative angle $\th$, the accumulated phases $\phi_F$ and $\phi_B$ are no longer equal, which results in a net spin current. This mechanism underlies the conventional ESC observed in junctions between AMs separated by an NM.

\begin{figure}[htb]
\includegraphics[width=4cm]{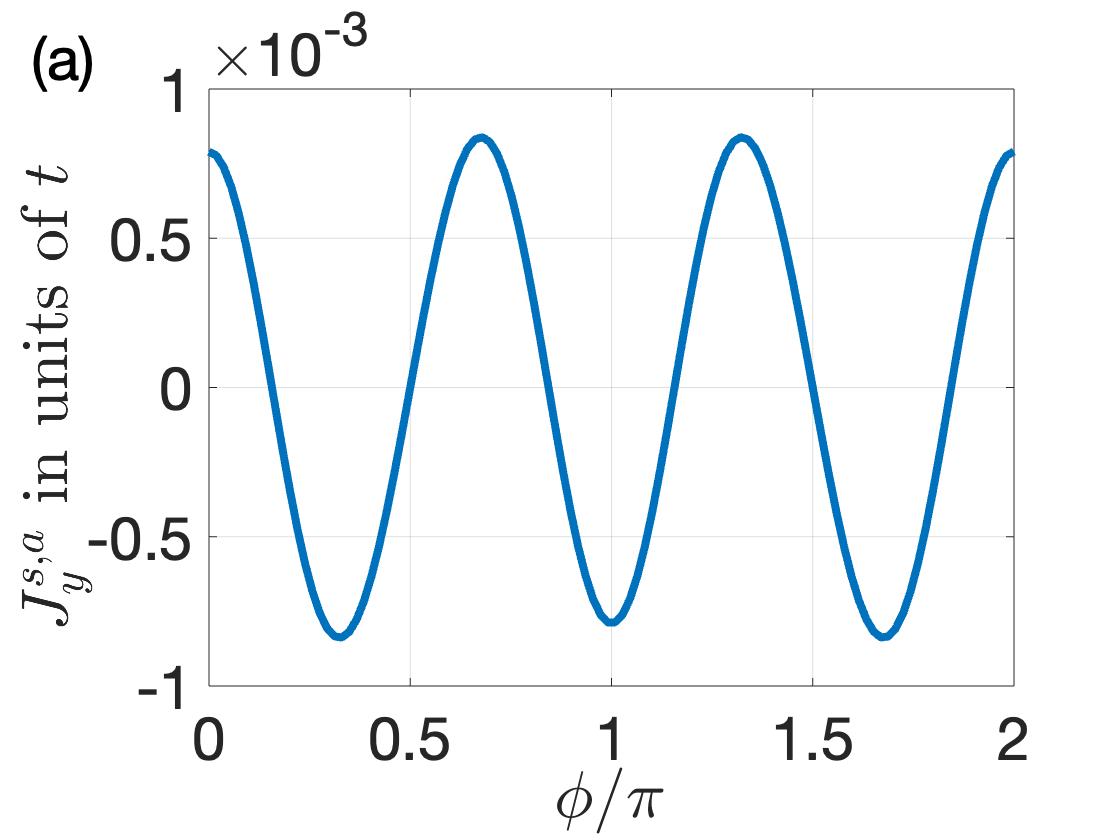}
\includegraphics[width=4cm]{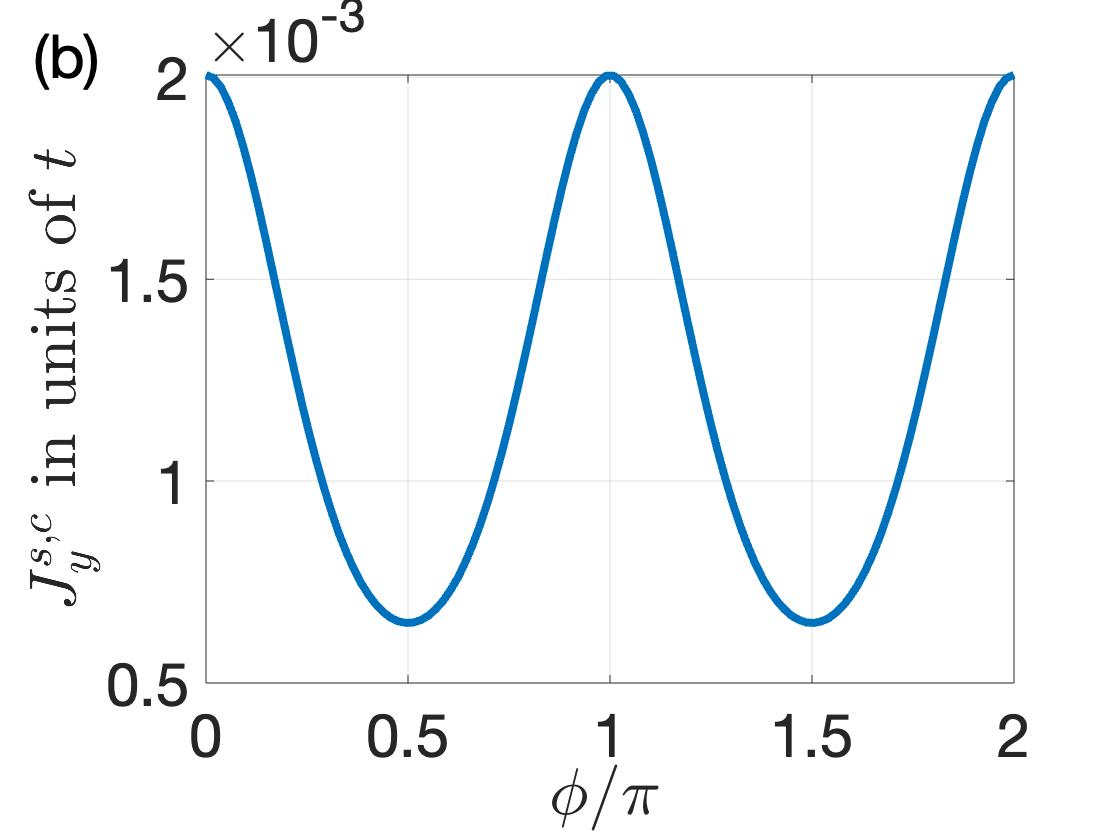}
\caption{(a) Anomalous ESC versus $\phi$ the angle made by the crystallographic axis along which the band bottoms for the two spin sectors are separated in PM with $x$-direction (b)~Critical current versus $\phi$. Parameters are same as in Fig.~\ref{fig:cpr-ano}, except $f=0.25$.}\label{fig:aesc}
\end{figure}

In Fig.~\ref{fig:aesc}, we plot (a) the anomalous ESC as a function of $\phi$, the angle formed by the crystallographic axis of the PM that separates the two band bottoms, and (b) the critical current versus $\phi$. We observe that both the anomalous ESC and the critical ESC, $J_y^{s,c}$, exhibit the symmetry property $J_y^{s,a/c}(\phi)=J_y^{s,a/c}(2\pi-\phi)$. This behavior can be understood by analyzing the accumulated phases $\phi_F$ and $\phi_B$, whose expressions remain invariant under the transformation $\phi \to 2\pi-\phi$.

Interestingly, the critical ESC possesses an additional symmetry: $J_y^{s,c}(\phi)=J_y^{s,c}(\pi-\phi)$. This arises because $\eta \to -\eta$ under the transformation $\phi\to \pi-\phi$. Consequently, the phases $\phi_F$ and $\phi_B$ interchange under the combined transformation $(\phi,\th)\to(\pi-\phi,-\th)$, which leads to a reversal of the spin current, $J_y^s\to -J_y^s$. However, due to the absence of a diode-like effect in the ESC, the magnitude of the critical current remains unchanged.

Furthermore, the critical current exhibits minima at $\phi=\pi/2$ and $3\pi/2$. The physical origin of the ESC lies in the reflection of an up-spin electron as a down-spin electron, and vice versa. These spin-flip reflection processes occur when electrons of opposite spins share the same transverse momentum, $k_y$. As $\phi$ increases from $0$ to $\pi/2$, the number of available states in the two spin bands with identical $k_y$ decreases, ultimately reaching a minimum at $\phi=\pi/2$.

\begin{figure}
\includegraphics[width=8cm]{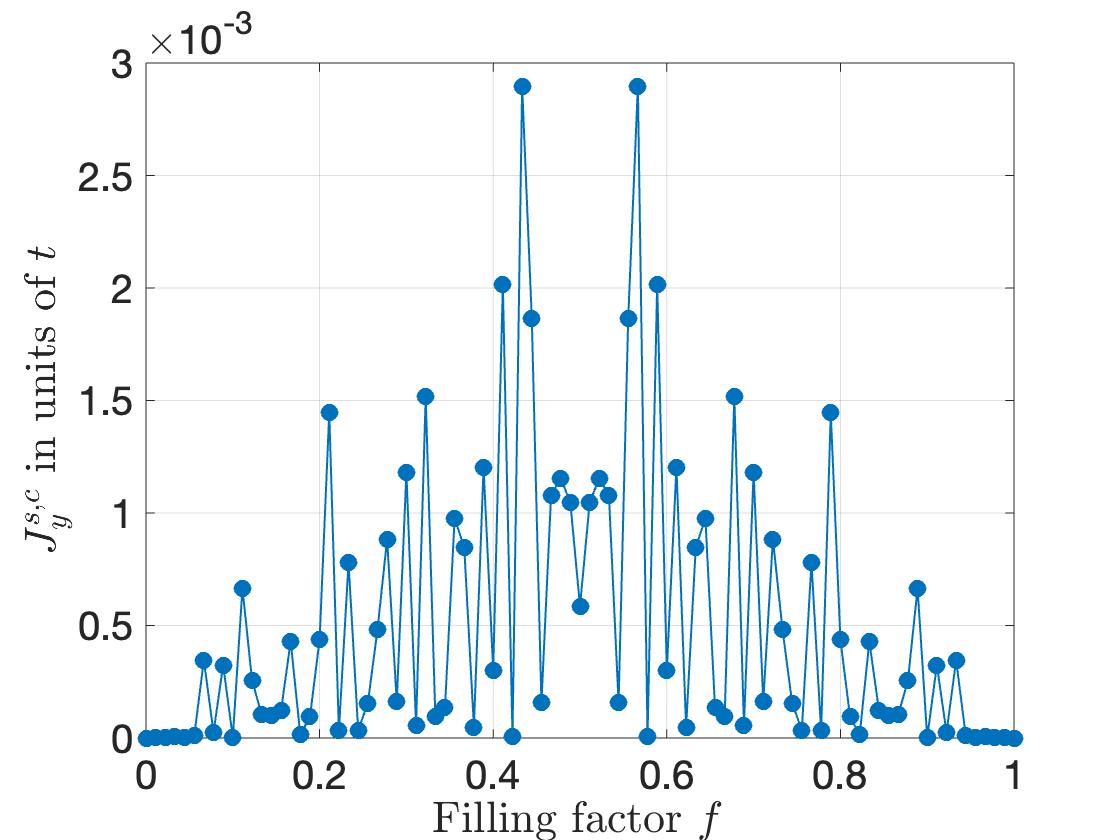}
\caption{Critical ESC versus filling factor. Parameters are same as in Fig.~\ref{fig:cpr1}.}\label{fig:Jyc-f}
\end{figure}
We note that the critical ESC exhibits pronounced, irregular fluctuations as a function of the filling factor $f$ [see Fig.~\ref{fig:Jyc-f}]. These rapid oscillations are fundamentally driven by mesoscopic quantum size effects and interference phenomena inherent to the finite geometry of the tight-binding lattice. Because the system has a finite transverse width ($L_y$), the transverse momentum $k_y$ is strictly quantized into discrete 1D subbands. As the filling factor changes, the tuning of the Fermi energy sweeps it past the bottoms of these discrete subbands. Each crossing corresponds to a van Hove singularity in the density of states and the abrupt opening of a new conduction channel. Furthermore, the finite central region of length $L_p$ acts as a quantum cavity. Electrons traversing this region accumulate a phase proportional to the Fermi momentum, leading to Fabry-P\'erot-like longitudinal resonances that change rapidly with changes in the energy. Finally, because the non-zero ESC heavily relies on spin-flip reflection processes that strictly require momentum matching ($k_y$) between opposite spins, the discrete nature of the transverse momentum causes the available phase space for these reflections to vary discontinuously. The complex interplay of transverse subband quantization, longitudinal cavity resonances, and discontinuous momentum-matching conditions collectively manifests as the strong fluctuations observed in the spin current.

We also note that the equilibrium spin current is symmetric about half-filling. This symmetry arises because the single-particle spin currents at energies $E$ and $-E$ are equal in magnitude but opposite in sign. Specifically, applying alternating minus signs to the real-space wavefunction amplitudes of an eigenstate at energy $E$ yields an eigenstate at energy $-E$. Because the expression for the local spin current between adjacent sites involves the product of their respective wavefunctions, this alternating-sign transformation exactly reverses the sign of the spin current. Consequently, since a completely filled band carries zero net current, the total spin current carried by the occupied states up to a filling factor of $(1-f)$ is equal to the negative of the spin current that would be carried by the unfilled states above the Fermi energy. Due to the $E \to -E$ symmetry of the single-particle currents, the current carried by these unfilled, high-energy states at filling $(1-f)$ is exactly the negative of the current carried by the low-energy occupied states up to filling $f$. This double-negative cancellation ensures that the macroscopic spin current takes the exact same value at filling factors $f$ and $(1-f)$.

\section{Summary and Conclusion}
\label{sec:conclusion}
In summary, we have investigated the emergence and behavior of equilibrium spin currents  in junctions comprising altermagnets   separated by either a normal metal   or a $p$-wave magnet   using a tight-binding lattice model. Because spin remains a good quantum number in the central scattering region, the equilibrium spin current is a well-defined and conserved quantity. By varying the relative angle $\th$ between the N\'eel vectors of the two AMs, we demonstrated that these magnetic junctions exhibit transport behaviors strongly analogous to the Josephson effect in superconducting junctions.

In the AM-NM-AM configuration, we found that the $\si_y$-polarized ESC exhibits a characteristic Josephson-like current-phase relation as a function of the relative angle $\th$. Crucially, this conventional ESC requires a misalignment between the N\'eel vectors and strictly vanishes when the AMs are parallel ($\th = 0$).

By replacing the central normal metal with a $p$-wave magnet, we explored the transport signatures of inversion symmetry breaking in an AM-PM-AM junction. Analogous to the simultaneous breaking of time-reversal and inversion symmetries required for the anomalous Josephson effect, the inversion-breaking nature of the PM drives an anomalous ESC. This anomalous spin current remains manifestly non-zero even when the N\'eel vectors of the two AMs are perfectly aligned ($\th = 0$). We showed that this phenomenon is fundamentally rooted in an asymmetry in the quantum phases ($\phi_F$ and $\phi_B$) accumulated by right- and left-moving electrons undergoing spin-flip reflections at the junction interfaces. 

Furthermore, we analyzed the dependence of both the anomalous ESC and the critical spin current on $\phi$, the angle characterizing the crystallographic axis of the PM. While both currents exhibit the symmetry $J_y^{s,a/c}(\phi)=J_y^{s,a/c}(2\pi-\phi)$, the critical current possesses an additional symmetry $J_y^{s,c}(\phi)=J_y^{s,c}(\pi-\phi)$ and exhibits pronounced minima at $\phi = \pi/2$ and $3\pi/2$. We attribute these minima to a reduction in the available transverse momentum ($k_y$) states that satisfy the stringent momentum-matching conditions required for the necessary spin-flip reflection processes.

Finally, we observed and explained the presence of rapid, irregular fluctuations in the critical ESC as a function of the band filling factor. These fluctuations are not artifacts, but physical manifestations of mesoscopic quantum size effects—specifically, the interplay between transverse subband quantization and longitudinal Fabry-P\'erot cavity resonances inherent to the finite geometry of the device.

Overall, our results establish that altermagnet junctions provide a rich platform for exploring dissipationless spin transport. The realization of both conventional and anomalous equilibrium spin currents in these systems paves the way for novel spintronic device paradigms that do not rely on charge transport or external biases.

\acknowledgements
The author thanks  Anusandhan National Research Foundation (ANRF, erstwhile SERB)   Core Research grant (CRG/2022/004311) and University of Hyderabad for financial support.  
\bibliography{ref_almag}
\end{document}